\documentclass{desyproc}

\usepackage{xspace}

\newcommand{\snn}{\ensuremath{\sqrt{s_{NN}}}\xspace}
\newcommand{\Raa}{\ensuremath{R_\mathrm{AA}}\xspace}
\newcommand{\Rdau}{\ensuremath{R_\mathrm{dAu}}\xspace}

\newcommand{\Npart}{\ensuremath{N_\mathrm{part}}\xspace}

\newcommand{\Jpsi}{\ensuremath{\mathrm{J}/\psi}\xspace}
\newcommand{\Ups}{\ensuremath{\Upsilon}\xspace}

\newcommand{\pT}{\ensuremath{p_\mathrm{T}}\xspace}

\newcommand{\Bee}{B_\mathrm{ee}\xspace}
\newcommand{\ee}{\mathrm{e}^{-}\mathrm{e}^{+}\xspace}

\begin{document}
%------------------------------------
\title{Upsilon Production at the STAR Experiment 
\\ \Large{with a Focus on New U+U Results} }

%for single authors the superscripts are optional
\author{{\slshape R\'obert V\'ertesi$^1$ for the STAR Collaboration}\\[1ex]
$^1$Nuclear Physics Institute ASCR, 25 068 \v{R}e\v{z}, Czech Republic}

% please enter the contribution ID for the DOI
\contribID{91}

% TO THE CONFERENCE EDITORS: 
% please update the following information      
% before sending the template to the authors
\confID{8648}  % if the conference is on Indico uncomment this line
\desyproc{DESY-PROC-2014-04}
\acronym{PANIC14} % if you want the Acronym in the page footer uncomment this line
\doi  % if there is an online version we will register DOIs

\maketitle

\begin{abstract}

We report recent \Ups measurements in p+p,
d+Au and Au+Au collisions at $\sqrt{s_{NN}}=200$ GeV, and detail the analysis in U+U
collisions at $\sqrt{s_{NN}}=193$ GeV.
Results on \Ups production versus rapidity are consistent with pQCD predictions in p+p collisions. 
However, \Ups production in mid-rapidity
  ($|y|${}$<$0.5) d+Au collisions is suppressed with respect to p+p
  collisions beyond model predictions that take into account
  modification of parton distribution functions and initial parton
  energy loss inside nuclei. The nuclear modification
factor \Raa shows a significant suppression in central Au+Au and
  U+U collisions, consistent with model calculations including color
screening effects in a deconfined medium. 
\end{abstract}

\section{Introduction}

Due to color screening, the production of quarkonia in high energy heavy
ion collisions is expected to be sensitive to the energy density of the
medium. Sequential suppression of different quarkonium states may
therefore serve as a thermometer of the medium~\cite{Mocsy:2007jz}. Although the suppression
of charmonia was anticipated as a key signature of the Quark Gluon
Plasma (QGP), the observed
energy dependence of $J/\psi$ suppression is rather weak~\cite{Adare:2006ns}. This
phenomenon is explained by $J/\psi$ production via recombination (coalescence) of $c\bar{c}$
pairs in the QGP. Bottomonia, on the other hand, are less affected by recombination
and can provide a cleaner probe of the strongly interacting medium.
While p+p measurements provide a benchmark for pQCD and serve as
a baseline for nuclear modification, d+Au collisions are generally
considered as suitable to study cold nuclear
matter (CNM) effects such as shadowing of the parton distribution functions
and initial state parton energy loss. Central U+U data at $\sqrt{s_{NN}}=193$ GeV, 
which is estimated to have a 20\% higher average energy density than that
of Au+Au~\cite{Kikola:2011zz}, allow for further tests of the sequential suppression
hypothesis. 

\section{Experiment and analysis}

\begin{figure}[h]
\begin{minipage}[t]{0.48\textwidth}
\centering
\includegraphics[trim=0 0 300 25, clip, width=0.9\columnwidth]{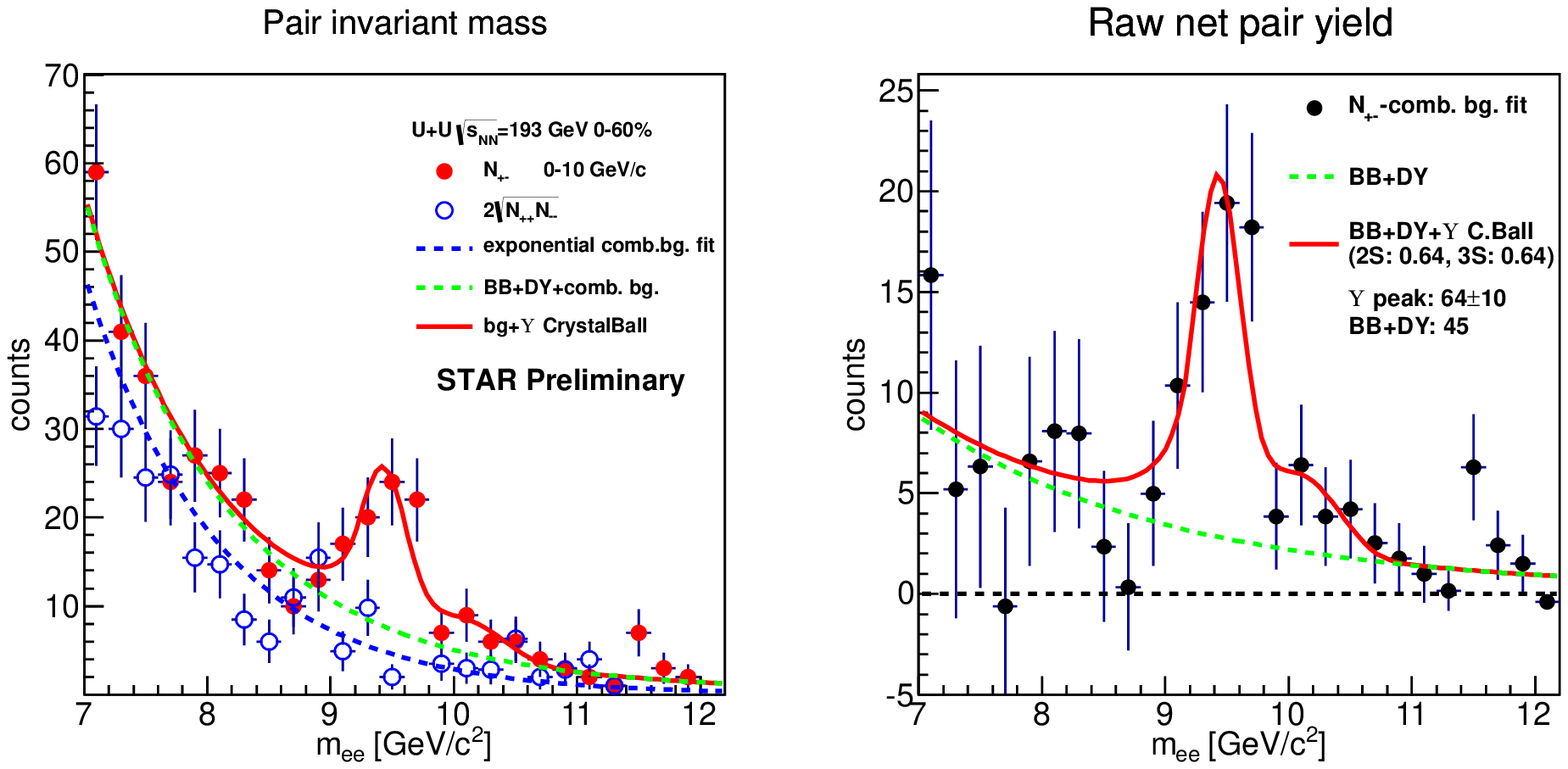}
\vspace{-1mm}
\caption{Invariant mass distribution of like-sign (filled dots) and unlike-sign
(open points) electron pairs in \snn{}=193 GeV U+U collisions of
0--60\% centrality at mid-rapidity, with background and peak fits.} 
\label{Fig:UpsInvM}
\end{minipage}\hspace{0.04\textwidth}%
\begin{minipage}[t]{0.48\textwidth}
\centering
\includegraphics[trim=0 0 0 30, clip=true, width=0.95\columnwidth]{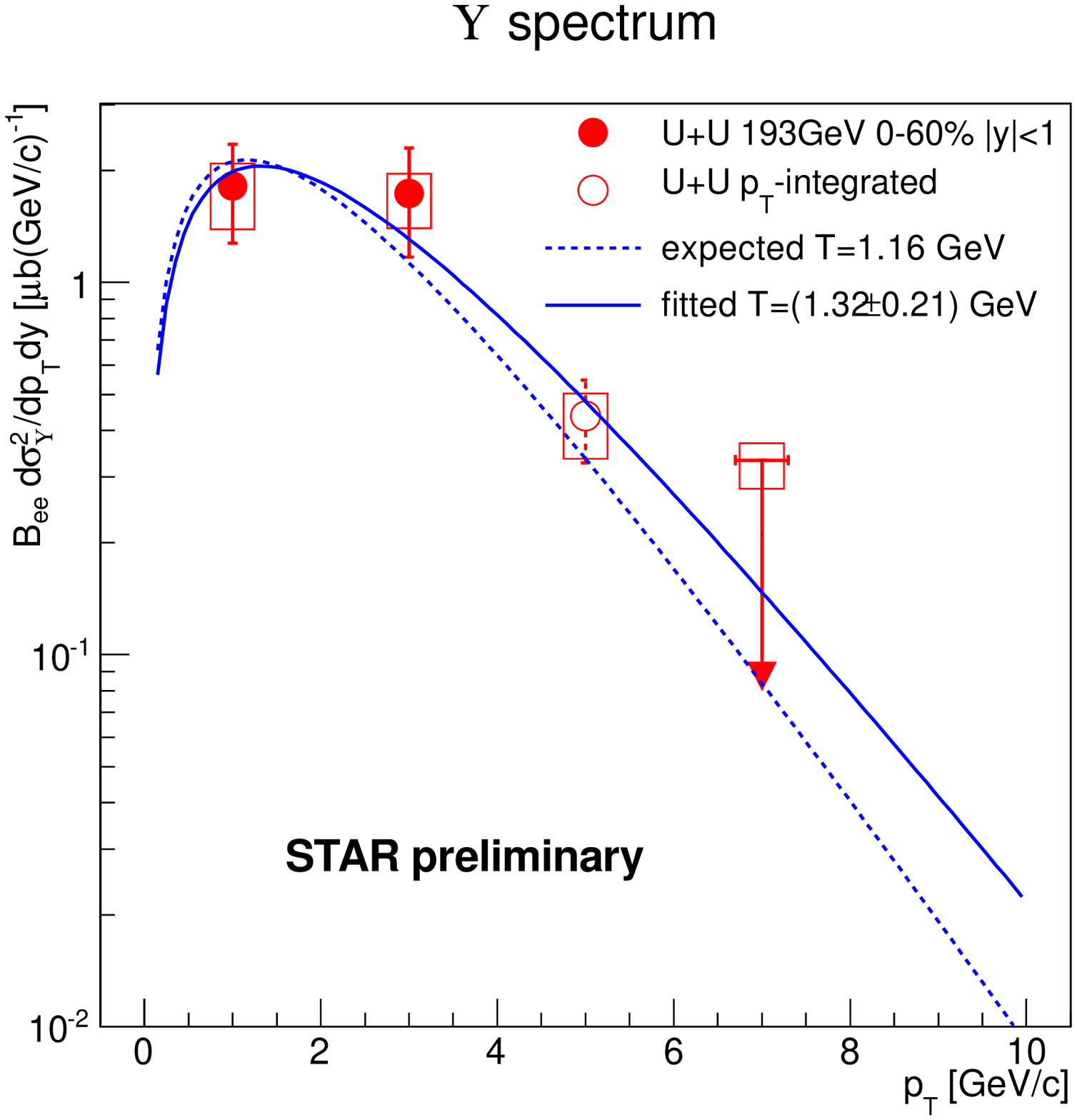}
\vspace{-1mm}
\caption{$\Upsilon$ \pT{}-spectrum in \snn{}=193 GeV U+U collisions of
0--60\% centrality at mid-rapidity. The fit (solid
line) to the data is compared to the expected slope (dashed).}
\label{Fig:UpsPtSpect}
\end{minipage} 
\end{figure}

The STAR experiment at RHIC is a complex detector that provides a full azimuthal coverage at 
mid-rapidity ($|\eta|${}$<$1). A detailed description of the STAR
  detector is in Ref.~\cite{Ackermann:2002ad}. 
The $\Ups\rightarrow \ee$ decay channel, with a
branching ratio $\Bee\approx2.4$ \%, was studied. 
Analysis of year 2012 $\sqrt{s_{NN}}=193$ GeV U+U data was done in a
similar way to recently published $\Upsilon$ measurements in p+p,
  d+Au and Au+Au collisions at $\sqrt{s_{NN}}=200$ GeV~\cite{Adamczyk:2013poh}, with differences highlighted below.
A total of 17.2 million {\it high-tower} triggered U+U events were
collected requiring an energetic hit in the Barrel Electromagnetic
Calorimeter (BEMC), corresponding to an integrated luminosity of 263.4 $\mu b^{-1}$.
Momentum measurement and electron identification
based on the energy loss $\mathrm{d}E/dx$ were done in the
Time Projection Chamber (TPC).
The projected position of the track is required to match the position of the
hit in the BEMC to the extent $\Delta
R=(\Delta\varphi^2+\Delta\eta^2)^{1/2}<0.04$ in the
azimuth--pseudorapidity space. 
The three most energetic adjacent BEMC towers including the hit tower  
were combined into {\it cluster}s.
Electron candidates were required to have similar cluster energy and momentum
($0.75<E_{cluster}/p<1.35\ c$) with most of the energy in one
tower ($E_{tower}/E_{cluster}>0.7$
for those candidates that fired the trigger, $E_{tower}/E_{cluster}>0.5$ for other
candidates). They were then paired, and required to have an
opening angle $\theta>90^\circ$. Fig.~\ref{Fig:UpsInvM} shows the invariant mass
distribution of the paired candidates. The combinatorial background was subtracted using
like-sign combinations. In the peak region there is also a significant
contribution from Drell-Yan and open $b\bar{b}$ processes. Templates of
the $\Ups(nS)$ peaks and the Drell-Yan contributions obtained from
simulations, and the $b\bar{b}$ contribution from pQCD model calculations were
fitted simultaneously to determine their relative contributions. The reconstruction
efficiency was determined using simulations and electron-enriched
  data samples as $\epsilon\approx 3\%$.
The corrected \pT-spectrum is shown in Fig.~\ref{Fig:UpsPtSpect}. Bin-shift correction was done using a
Boltzmann function with a slope $T=1.16$ GeV, extracted from a
parametrized interpolation over ISR, CDF and CMS data. 
A fit to the spectrum yiels a slope $T=1.32\pm0.21$ GeV, consistent
with the interpolation. 
The measured \Ups cross section in U+U collisions is $\Bee
\left. \frac{d\sigma^\Upsilon_{\rm{AA}}}{dy}\right|_{|y|<1} = 4.37 \pm
1.09 (stat) \pm ^{+0.65}_{-1.01} (syst)\ \mu\rm{b}$.
The major systematic
uncertainties are from signal extraction ($^{+4.8}_{-18}$\%), tracking
  efficiency (11.8\%), electron identification in the TPC
  ($^{+4.0}_{-6.4}$\%) and in the BEMC (5.9\%), TPC-BEMC
    matching (5.4\%), trigger efficiency ($^{+1.1}_{-3.6}$\%), geometrical
      acceptance  ($^{+1.7}_{-3.0}$\%) and input \pT and $y$ spectrum in the simulations (2.1\%). 

\section{Upsilon production in p+p and d+Au collisions}

\begin{figure}[h]
\begin{minipage}[t]{0.48\textwidth}
\centering
\includegraphics[width=0.9\columnwidth]{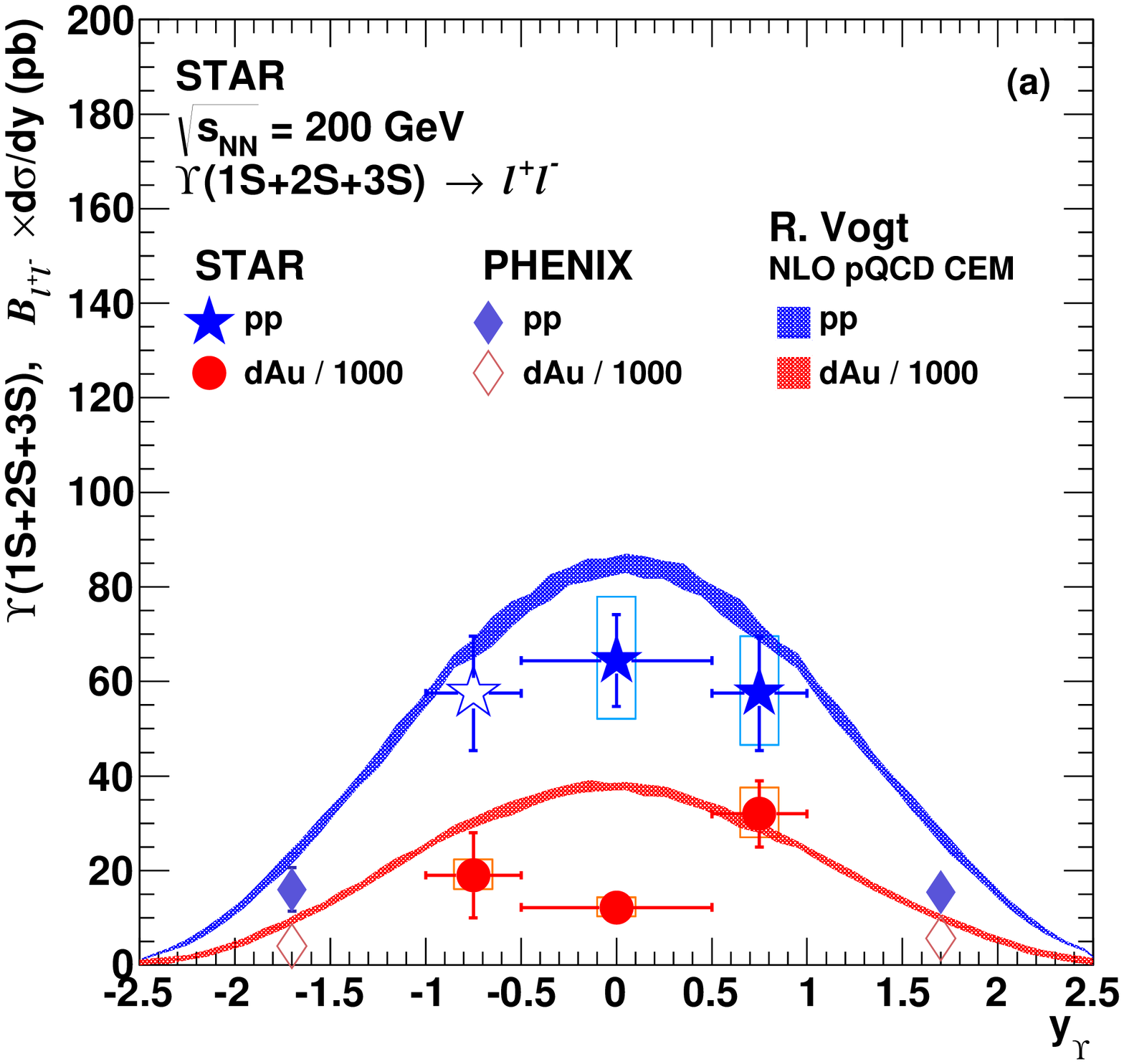}
\vspace{-1mm}
\caption{\Ups yield $\Bee \frac{d\sigma}{dy}$ for p+p and d+Au collisions~\cite{Adamczyk:2013poh}
compared to a pQCD model~\cite{Vogt:2012fba}.
} 
\label{Fig:UpsXsectY}
\end{minipage}\hspace{0.04\textwidth}%
\begin{minipage}[t]{0.48\textwidth}
\centering
\includegraphics[width=0.87\columnwidth]{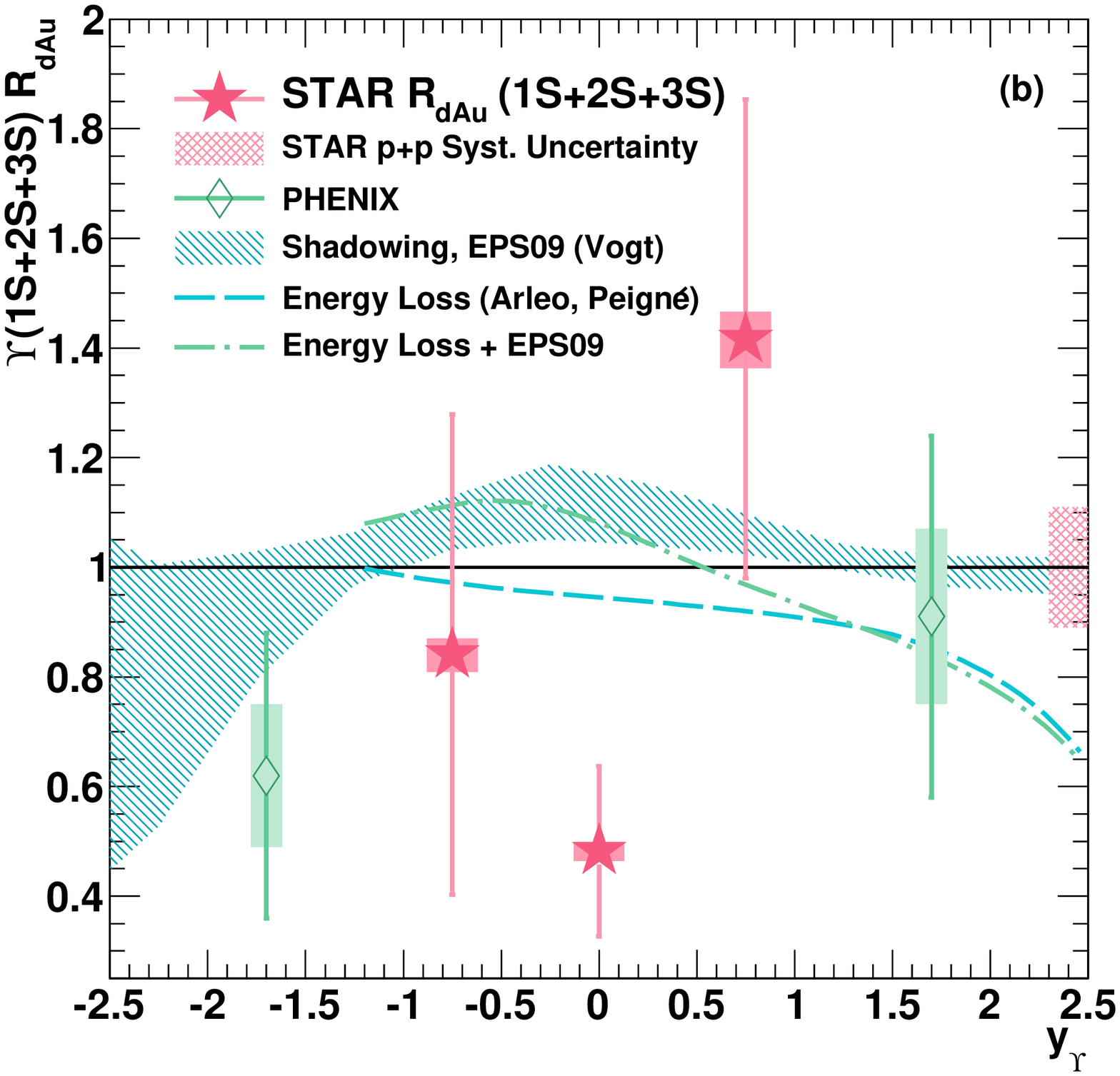}
\vspace{-1mm}
\caption{\Rdau versus $y$ of \Ups{} mesons~\cite{Adamczyk:2013poh} compared to 
theoretical calculations~\cite{Vogt:2012fba,Arleo:2012rs}.}
\label{Fig:UpsRaaY}
\end{minipage} 
\end{figure}

Fig.~\ref{Fig:UpsXsectY} shows the cross sections for \Ups
production in \snn{}=200 GeV p+p and d+Au collisions~\cite{Adamczyk:2013poh}. The data are
compared to NLO pQCD color evaporation model
predictions~\cite{Vogt:2012fba}. In Fig.~\ref{Fig:UpsRaaY} the nuclear modification factor in
d+Au is compared to calculations including shadowing and/or parton energy loss~\cite{Vogt:2012fba, Arleo:2012rs}. While the p+p data are consistent with pQCD,
CNM effects alone may not be enough to explain the suppression in the
d+Au mid-rapidity bin ($|y|${}$<$0.5). 

\section{Upsilon suppression in heavy ion collisions}

\begin{figure}[h]
\begin{minipage}[b]{0.48\textwidth}
\centering
\includegraphics[width=0.9\columnwidth]{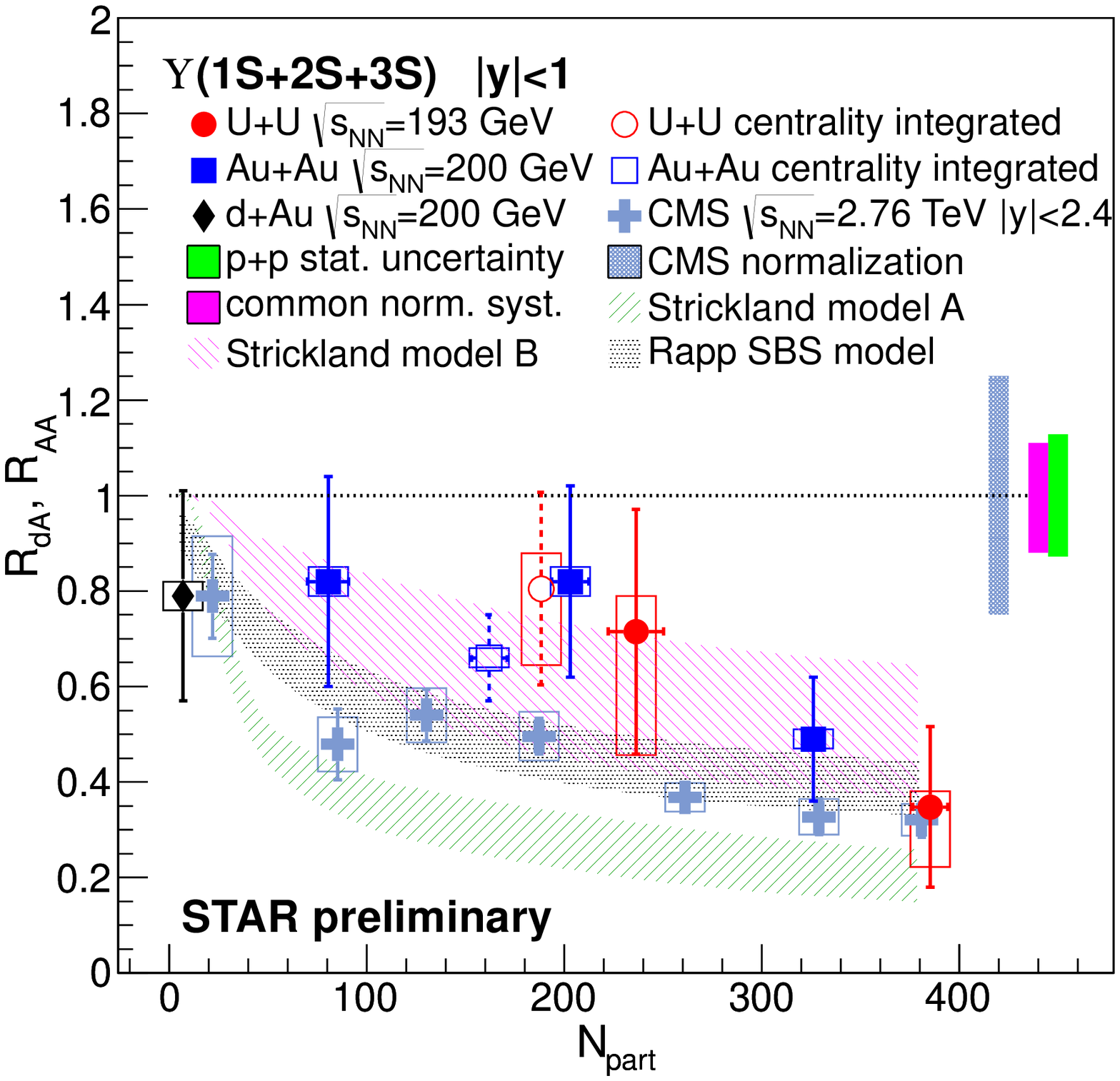}
\vspace{-3mm}
\caption{\Raa{} vs. \Npart in \snn{}=200 GeV d+Au, Au+Au and 193 GeV U+U
collisions, compared to models~\cite{Strickland:2011aa,Emerick:2011xu} and
LHC data~\cite{Chatrchyan:2012lxa}.
} 
\label{Fig:UpsRaaNpart}
\end{minipage}\hspace{0.04\textwidth}%
\begin{minipage}[b]{0.48\textwidth}
\centering
\includegraphics[width=\columnwidth]{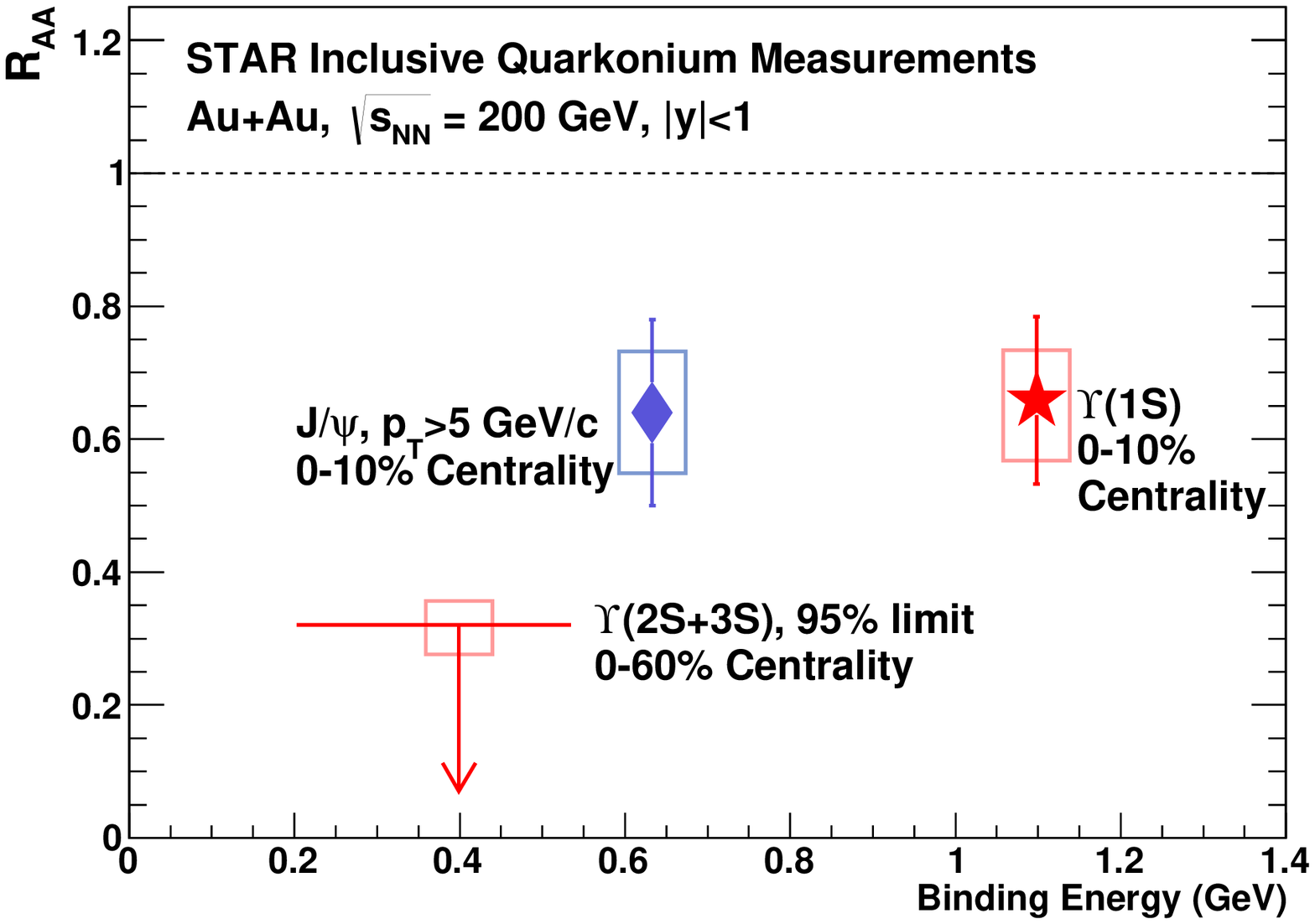}
\caption{\Raa of the $\Upsilon$(1S) and the $\Upsilon$(2S+3S) states compared to high-\pT
\Jpsi \Raa~\cite{Adamczyk:2012ey}, plotted against binding energy, in \snn{}=200 GeV Au+Au collisions.}
\label{Fig:UpsRaaStates}
\end{minipage} 
\end{figure}

Nuclear modification factors of the $\Upsilon$(1S+2S+3S) in d+Au, Au+Au and U+U
collisions are presented in Fig.~\ref{Fig:UpsRaaNpart} with respect to the number
of participants, and compared to model calculations~\cite{Strickland:2011aa,Emerick:2011xu}, as well as \snn{}=2.76
TeV Pb+Pb data from the CMS experiment~\cite{Chatrchyan:2012lxa}. The
trend observed in Au+Au is generally continued in the U+U data, with an $R_{\rm{AA}}=0.35\pm 0.17 (stat.) ^{+0.03}_{-0.13}
(syst.)$ in the 10\% most central U+U collisions. 
The model of Strickland and Bazow~\cite{Strickland:2011aa} incorporates lattice QCD
results on screening and broadening of bottomonium, as well as the dynamical
propagation of the $\Upsilon$ meson in the colored medium. The scenario
with a potential based on heavy quark internal energy is consistent
with the observations, while the free energy based scenario is
disfavoured. The strong binding scenario in a model proposed by Emerick,
Zhao, and Rapp~\cite{Emerick:2011xu}, which includes possible CNM effects in
addition, is also consistent with STAR results. 
The measured \Raa at RHIC and at LHC are consistent within the
sizeable uncertainties. However, the LHC data, which corresponds to
  higher energy densities, shows a trend that differs from RHIC: a
  strong suppression is present at all but the lowest \Npart values.

Fig.~\ref{Fig:UpsRaaStates} shows Au+Au \Raa for the ground state $\Upsilon$(1S) and the
excited states $\Upsilon$(2S+3S) separately, compared to the \Raa of
high-\pT \Jpsi mesons in $\sqrt{s_{NN}}=200$ GeV Au+Au collisions~\cite{Adamczyk:2012ey}.
The  $\Upsilon$(1S) shows a suppression similar to that of high-\pT
\Jpsi mesons, more than if only cold
nuclear matter effects were present ~\cite{Adamczyk:2013poh}. The excited state yields are
consistent with a complete suppression within the precision of the measurement. 

\section{Summary and outlook}

We reported recent measurements of \Ups production in p+p, d+Au and
Au+A collisions at $\sqrt{s_{NN}}=200$ GeV, as well as U+U at 193 GeV. The slope of the \pT-spectrum in U+U collisions is consistent with interpolations from other experiments. We see a significant suppression
in $|y|${}$<$1 central Au+Au and U+U collisions, which attests to the
presence of a deconfined medium and support the sequential melting hypothesis.
However, the $|y|${}$<$0.5 d+Au data also shows a suppression beyond model
predictions, suggesting that CNM effects may also play an
important role. 
The new Muon Telescope Detector has been completed by 2014, and
will allow for a precise reconstruction of the three \Ups states
separately, through the dimuon channel. Future high-statistics p+Au collisions from 2015 will
help us gain a deeper insight to the CNM effects.

This work has been supported by the grant 13-02841S of the Czech
Science Foundation (GA\v{C}R), and by the MSMT grant CZ.1.07/2.3.00/20.0207 of the European Social Fund (ESF) in the Czech Republic: “Education for Competitiveness Operational Programme” (ECOP).

\begin{footnotesize}

\end{footnotesize}

\end{document}